%% file: is2021_e2e_fusion_v2.0_final_fixed.tex
\documentclass[a4paper]{article}
\usepackage{hhline}
\usepackage{multirow}
\usepackage[table]{xcolor}
\usepackage{colortbl}
\usepackage{INTERSPEECH2020}
\usepackage{ifthen}
\usepackage{array}

\input{macros/tikz_init}
\usetikzlibrary{arrows.meta}
\input{macros/mathmacros}

\ninept

\let\OLDthebibliography\thebibliography
\renewcommand\thebibliography[1]{
	\OLDthebibliography{#1}
	\setlength{\parskip}{0pt}
	\setlength{\itemsep}{0.7pt plus 0.3ex}
}

\pgfplotsset{compat=newest}

\makeatletter
\newcommand{\gettikzxy}[3]{%
  \tikz@scan@one@point\pgfutil@firstofone#1\relax
  \edef#2{\the\pgf@x}%
  \edef#3{\the\pgf@y}%
}
\makeatother

\title{Multi-Encoder Learning and Stream Fusion for \\ Transformer-Based End-to-End Automatic Speech Recognition}
\name{Timo Lohrenz, Zhengyang Li, Tim Fingscheidt}
\address{Technische Universit\"at Braunschweig\\
	Institute for Communications Technology\\ 
	Schleinitzstr. 22, 38106 Braunschweig, Germany
	}
\email{t.lohrenz@tu-bs.de, zhengyang.li@tu-bs.de, t.fingscheidt@tu-bs.de}

\begin{document}

\maketitle

\begin{abstract}\vspace{-1mm}
Stream fusion, also known as system combination, is a common technique in automatic speech recognition for traditional hybrid hidden Markov model approaches, yet mostly unexplored for modern deep neural network end-to-end model architectures. Here, we investigate various fusion techniques for the all-attention-based encoder-decoder architecture known as the transformer, striving to achieve optimal fusion by investigating different fusion levels in an example single-microphone setting with fusion of standard magnitude and phase features. We introduce a novel multi-encoder learning method that performs a weighted combination of two encoder-decoder multi-head attention outputs \textit{only} during training. Employing then only the magnitude feature encoder in inference, we are able to show consistent improvement on Wall Street Journal (WSJ) with language model and on Librispeech, without increase in runtime or parameters. Combining two such multi-encoder trained models by a simple late fusion in inference, we achieve state-of-the-art performance for transformer-based models on WSJ with a significant WER reduction of 19\% relative compared to the current benchmark approach.

%Employing then only the magnitude feature encoder in inference, the WER of standard transformer approaches decreases by up to 0.8\% absolute across all investigated tasks without increasing the amount of parameters during inference.

\end{abstract}

\noindent\textbf{Index Terms}: End-to-end speech recognition, information fusion, multi-encoder learning, transformer, phase features

\section{Introduction}

%% Hybrid to end2end ASR 
In recent years a paradigm shift in automatic speech recognition (ASR) research is seen towards the replacement of established hybrid hidden Markov model (HMM) based approaches~\cite{bourlard1994} by end-to-end trained neural networks making pronunciation dictionaries and phonetic modeling techniques obsolete~\cite{bahdanau2016}. Proposed methods for end-to-end training include connectionist temporal classification (CTC)~\cite{Graves2014}, recurrent neural network transducers (RNN-T)~\cite{Graves2013} and recent attention-based encoder-decoder (AED) models, namely the listen-attend-and-spell (LAS)~\cite{chan2016} and the transformer model~\cite{Vaswani2017}. While the LAS models employ recurrent connections in the typical encoder-decoder structure of the end-to-end models, transformer models rely entirely on the attention mechanism to capture temporarily relevant information in speech~\cite{Dong2018c}. On large datasets such as Librispeech~\cite{Panayotov2015}, transformer models outperform hybrid speech recognition already by a large margin~\cite{Karita2019}.

%% Techniques for hybrid ASR transferred to End2End, Fokus auf Fusion
For the well-established hybrid speech recognition stream fusion approaches can be classified into three categories based on which stage in the system fusion is performed: early fusion---combination in the input feature domain~\cite{Potamianos2001,Schlueter2006,Ravanelli2019}, middle fusion---combination of an intermediate information representation~\cite{Luettin2001,Misra2003,lohrenz2020} (e.g.,\ state likelihoods), or late fusion---combination of system outputs (e.g.,\ word hypotheses~\cite{Fiscus1997}, output posteriors, confusion networks~\cite{Hoffmeister2006}, or lattices~\cite{xu2011}). A prominent task for fusion is audiovisual automatic speech recognition (AV-ASR)~\cite{Potamianos2003,Receveur2016}, employing additional visual sensors to increase robustness in noisy conditions. In single-channel settings, usually different feature representations are used for fusion (e.g.,\ filterbank and fMLLR features~\cite{Ravanelli2019}, different short-time Fourier transform window sizes~\cite{lohrenz2020}, or as here, standard magnitude features with phase features~\cite{Lohrenz2017}), or multiple acoustic models~\cite{Du2018,Xiong2017}.

%% Audiovisual Speech Recognition and other fusion settings:
Concerning the fusion of additional information into end-to-end transformer models, the few existing approaches stem from audiovisual automatic speech recognition~\cite{Petridis2018,Afouras2018} and neural machine translation~\cite{li2020c}, where additional encoders are used to gather visual speech information or contextual information, respectively. Recent successful non-fusion techniques for end-to-end models are multi-task learning, e.g., by using a combination of CTC and attention-based losses~\cite{kim2017,moriya2020}, and augmentation techniques such as spectral augmentation~\cite{Park2019d}. Those methods improve neural networks by adding more variety to the trained models either by composite losses or by randomly withholding information in the input features. An unexplored approach to add such variety, strongly related to fusion, is to use multiple encoders during training of the transformer model. Some context-aware approaches in neural machine translation use additional context encoders~\cite{Zhang2018g} to incorporate previous context of a sentence to achieve a better translation, while in~\cite{li2020c} it has been found that results improve even if such context is ignored during inference.

%% In this paper... 
In this paper we adopt and modify fusion techniques from hybrid ASR to transformer-based end-to-end speech recognition on an exemplary audio-only fusion task by combining the common magnitude-based feature representations with additional phase-based features. To elaborate the best possible fusion method we apply simple input feature and output posterior combination methods as well as middle fusion schemes, that use two different encoders to perform fusion of the respective encoder-decoder multi-head attention outputs. For this middle fusion approach we investigate several variants comprising an optional sharing of the encoder-decoder attention parameters as well as different paradigms combining the outputs thereof. In addition we explore a novel method which we dub \textit{multi-encoder learning} (MEL) that uses both individual encoders \textit{only} during training, thereby increasing robustness of the standard non-fusion transformer even during single encoder inference.

%% Structure
The paper is structured as follows: In Section 2, we introduce known and novel fusion and learning approaches to end-to-end model architectures. Section 3 describes the fusion experiment setup on Wall Street Journal (WSJ) and Librispeech, while corresponding results are reported and discussed in Section 4. The paper is concluded in Section 5.

\begin{figure}[t!]
	\centering
	\tikzset{%
		tipA/.tip={Triangle[angle=40:4pt]}
	}
	\tikzset{
		>=stealth',
		punktchain/.style={
			rectangle,rounded corners, 
			draw=black, thick,
			text width=7em, 
			minimum height=0.4cm, 
			text centered,fill=white, 
			on chain=a},
		line/.style={draw, thick, ->},
		every join/.style={->, thick,shorten >=1pt},
	}
	\tikzset{
		>=stealth',
		punkt_mult/.style={
			circle, 
			minimum size=0.1cm,inner sep=0mm, 
			draw=black, 
			text centered,fill=white,thick, 
			on chain},
		arrow/.style={draw, thick,-latex},
		every join/.style={->, thick,shorten >=1pt},
	}
	\tikzset{
	>=stealth',
	loet/.style={
		draw,circle,fill=black,inner sep=0.3mm,outer sep=0.0mm},
	arrow/.style={draw, thick,-tipA},
	every join/.style={->, thick,shorten >=1pt},
}
	% defining the new dimensions and parameters

	\begin{tikzpicture}[node distance=0.4cm,scale=1, every node/.style={scale=0.9}]
	
	{  [start chain=a going above]
		\def\x{-8mm}
		\def\sc{1.11}
		
		\node[on chain=a,outer sep=0mm] (feats) {$\mathbf{o}_1^T$};
		\node[punktchain,text width=2.1cm,yshift=-0.0cm,xshift=\x*\sc] (CONV) {$\mathrm{CNN}$};
		\node[punkt_mult,yshift=1mm,xshift=-\x*\sc] (enc_pos_emb) {$\mathbf{+}$};
		\node[draw,circle, left=of enc_pos_emb, xshift=2mm,inner sep= 0mm] (pos_enc) {\tikz \draw[x=0.95mm,y=1.3mm] (0,0) sin (1,1) cos (2,0) sin (3,-1) cos (4,0);};
		\node[left=of pos_enc,draw=none, yshift=0mm, text width=1.1cm,xshift=5mm,align=center] (pos_enc_label){\footnotesize positional \\ \vspace{-1mm} encoding};
 		\draw[arrow] ( pos_enc) -- (enc_pos_emb);

		\node[punktchain,draw=black,fill=green!0,xshift=\x*\sc,](Encoder1)      {Encoder Block};
		\node[on chain=a,xshift=\x*\sc](dots){...};
		% \node[right=of dots,xshift=-0.6cm](dots_label){\scriptsize +10 ENC Blocks};
		\node[right=of dots,xshift=-0.75cm, text width=1.8cm,align=center,yshift=1mm](dots_label){\scriptsize in total \\ \vspace{-1mm} 12 ENC blocks};
		
		\node[punktchain,draw=black,fill=green!0,xshift=-\x*\sc](Encoder2)      {Encoder Block};
		\node[left=of feats,draw=none, xshift=1.0cm,yshift=-0.25cm,text width=2.1cm](output_label_label)      {\footnotesize magnitude \\ \vspace{-1mm}  feature sequence};
		\node[on chain=a,draw=none,xshift=1.55cm](out_node)      {};
		
		\draw[arrow] ( feats.north) --node[left,text=black!80,pos=0.3] {\scriptsize$B  \times 1 \times T \times 83$} ([xshift=-\x] CONV.south);
		\draw[arrow] ( [xshift=-\x]CONV.north) --node[left,text=black!80,pos=0.3] {\scriptsize$B\times T/4 \times d$} (enc_pos_emb.south);
		\draw[arrow] ( enc_pos_emb.north) -- ([xshift=-\x] Encoder1.south);
		\draw[arrow] ([xshift=\x] Encoder1.north) --node[right,text=black!80,pos=0.3] {\scriptsize$B\times T/4 \times d$} ( dots.south);
		\draw[arrow] (dots.north) --node[left,text=black!50] {} ([xshift=\x] Encoder2.south);
		\draw[arrow,-] ([xshift=\x] Encoder2.north) |- node[right,text=black!80,pos=0.2] {\scriptsize$B\times T/4 \times d$} (out_node.south);
		
	}
	
	{  [start chain=b going above]
		\def\x{-8mm}
		\def\sc{1.11}
		
		\node[punktchain,on chain=b,right=of CONV,xshift=0.3cm,text width = 2.1cm] (embed) { Embedding};
		\node[draw=none,below=of embed,outer sep=0.5mm,yshift=0.0cm,xshift=-\x*\sc] (input_character) {${c}_{\ell\!-\!1}$};
		\node[punkt_mult,yshift=1mm,on chain=b,xshift=-\x*\sc] (enc_pos_emb) {$\mathbf{+}$};
		\node[draw,circle, left=of enc_pos_emb, xshift=2mm,inner sep= 0mm] (pos_enc) {\tikz \draw[x=0.95mm,y=1.3mm] (0,0) sin (1,1) cos (2,0) sin (3,-1) cos (4,0);};
		\node[left=of pos_enc,draw=none, yshift=0mm, text width=1.1cm,xshift=5mm,align=center] (pos_enc_label){\footnotesize positional \\ \vspace{-1mm} encoding};
		\draw[arrow] ( pos_enc) -- (enc_pos_emb);
		
		\node[punktchain,on chain=b,draw=black,fill=black!10,xshift=\x*\sc](Decoder1)      {Decoder Block};
		\node[on chain=b,xshift=\x*\sc](decoder_dots){...};
		%\node[right=of decoder_dots,xshift=-0.6cm](decoder_dots_label){\scriptsize +4 DEC Blocks};
		\node[right=of decoder_dots,xshift=-0.75cm, text width=1.8cm,align=center,yshift=1mm](dots2_label){\scriptsize in total \\ \vspace{-1mm} 6 DEC blocks};
		\node[punktchain,on chain=b,draw=black,fill=black!10,xshift=-\x*\sc](Decoder2)      {Decoder Block};

		\node[punktchain,on chain=b,draw=black,yshift=0.6cm, text width=2.1cm,,rounded corners=0pt](output_ln)      {Layer Norm};
		\node[punktchain,on chain=b,draw=black,text width=7em,rounded corners=0pt](output_fc)      {Fully Connected + Softmax};
		\node[on chain=b,draw=none,xshift=\x*\sc](output_label)      {$\VEC{P}_\ell$};
		\node[right=of output_label,draw=none, yshift=-0.0cm](output_label_label)      {\footnotesize output token probabilities};
		\node[left=of input_character,draw=none, xshift=1.4cm,yshift=-0.25cm,text width=2.1cm](output_label_label)      {\footnotesize previous \\ \vspace{-1mm} input character};
		
		\draw[arrow] (input_character.north) --node[left,text=black!80,pos=0.36] {\scriptsize$B \times 1 $} ([xshift=-\x]embed.south);
		\draw[arrow] ([xshift=-\x] embed.north) --node[left,text=black!80,pos=0.3] {\scriptsize$B \times 1 \times d$} (enc_pos_emb.south);
		\draw[arrow] ( enc_pos_emb.north) -- ([xshift=-\x] Decoder1.south);
		\draw[arrow] ([xshift=\x] Decoder1.north) --node[right,text=black!80,pos=0.3] {\scriptsize$B \times 1 \times d$} (decoder_dots.south);
		\draw[arrow] ( decoder_dots.north) --node[right,text=black!50] {} ([xshift=\x] Decoder2.south);
		\draw[arrow] ([xshift=\x]  Decoder2.north) --node[right,text=black!80] {} ([xshift=\x] output_ln.south);
		\draw[arrow] ([xshift=\x] output_ln.north) --node[right,text=black!80] {\scriptsize$B \times 1 \times d$} ([xshift=\x] output_fc.south);
		
		\draw[arrow] ([xshift=\x] output_fc.north) --node[right,text=black!80] {\scriptsize$B \times 1 \times D$} (output_label.south);
		
		% Löt-Dots
		\node[loet,left of=Decoder2,xshift=-1.13cm]	(dec_block_2_mag_in)      {};
		\node[loet,left of=decoder_dots,xshift=-0.25cm,yshift=1mm]	(dec_block_mag_dot_in1) {} ;
			\draw[arrow] (dec_block_mag_dot_in1.east) -- ++(2.5mm,0em);
		\node[loet,left of=decoder_dots,xshift=-0.25cm,yshift=3mm]	(dec_block_mag_dot_in2) {} ;
			\draw[arrow] (dec_block_mag_dot_in2.east) -- ++(2.5mm,0em);		
		\node[loet,left of=decoder_dots,xshift=-0.25cm,yshift=-1mm]	(dec_block_mag_dot_in3) {} ;
			\draw[arrow] (dec_block_mag_dot_in3.east) -- ++(2.5mm,0em);	
		\node[loet,left of=decoder_dots,xshift=-0.25cm,yshift=-3mm]	(dec_block_mag_dot_in3) {} ;
			\draw[arrow] (dec_block_mag_dot_in3.east) -- ++(2.5mm,0em);	
			
		\node[loet,right of=Decoder2,xshift=1.13cm]	(dec_block_2_phase_in)      {};
		\node[loet,right of=decoder_dots,xshift=2.01cm,yshift=1mm]	(dec_block_phase_dot_in1) {} ;
			\draw[arrow] (dec_block_phase_dot_in1.west) -- ++(-2.5mm,0em);
		\node[loet,right of=decoder_dots,xshift=2.01cm,yshift=3mm]	(dec_block_phase_dot_in2) {} ;
			\draw[arrow] (dec_block_phase_dot_in2.west) -- ++(-2.5mm,0em);	
		\node[loet,right of=decoder_dots,xshift=2.01cm,yshift=-1mm]	(dec_block_phase_dot_in3) {} ;
			\draw[arrow] (dec_block_phase_dot_in3.west) -- ++(-2.5mm,0em);	
		\node[loet,right of=decoder_dots,xshift=2.01cm,yshift=-3mm]	(dec_block_phase_dot_in4) {} ;
			\draw[arrow] (dec_block_phase_dot_in4.west) -- ++(-2.5mm,0em);	
	}

	\draw[|-,arrow] ( out_node) |- node[right,text=black!50] {} (Decoder1.west);
	\draw[|-,arrow] ( out_node) |- node[right,text=black!50] {} (Decoder2.west);
	
	{  [start chain=c going above]
		\def\x{-8mm}
		\def\sc{1.11}
		\node[punktchain,on chain=c,right=of embed, xshift=0.3cm,text width=2.1cm] (CONV2) {$\mathrm{CNN}$};
		\node[yshift=0.0cm,below=of CONV2,xshift=-\x*\sc] (feats2) {$\mathbf{u}_1^T$};
		
		\node[punkt_mult,yshift=1mm,on chain=c,xshift=-\x*\sc] (enc_pos_emb2) {$\mathbf{+}$};
		\node[draw,circle, left=of enc_pos_emb2, xshift=2mm,inner sep= 0mm] (pos_enc) {\tikz \draw[x=0.95mm,y=1.3mm] (0,0) sin (1,1) cos (2,0) sin (3,-1) cos (4,0);};
\node[left=of pos_enc,draw=none, yshift=0mm, text width=1.1cm,xshift=5mm,align=center] (pos_enc_label){\footnotesize positional \\ \vspace{-1mm} encoding};
\draw[arrow] ( pos_enc) -- (enc_pos_emb2);

		\node[punktchain,on chain=c,draw=black,fill=green!0,xshift=\x*\sc](Encoder12)      {Encoder Block};
		\node[on chain=c,xshift=\x*\sc](dots2){...};
		\node[right=of dots2,xshift=-0.75cm, text width=1.8cm,align=center,yshift=1mm](dots2_label){\scriptsize in total \\ \vspace{-1mm} 12 ENC blocks};
		\node[left=of feats2,draw=none, xshift=1.0cm,yshift=-0.25cm,text width=2.1cm](output_label_label)      {\footnotesize phase \\ \vspace{-1mm} feature sequence};
		\node[punktchain,on chain=c,draw=black,fill=green!0,xshift=-\x*\sc](Encoder22)      {Encoder Block};
		
		\node[on chain=c,draw=none,xshift=-1.55cm](out_node2)      {};
		
		\draw[arrow] ( feats2.north) --node[left,text=black!80,pos=0.3] {\scriptsize$B  \times 1 \times T \times 83$} ([xshift=-\x] CONV2.south);
		\draw[arrow] ([xshift=-\x] CONV2.north) --node[left,text=black!80,pos=0.3] {\scriptsize$B\times T/4 \times d$} ( enc_pos_emb2.south);
		\draw[arrow] ( enc_pos_emb2.north) -- ([xshift=-\x] Encoder12.south);
		\draw[arrow] ([xshift=\x] Encoder12.north) --node[right,text=black!80,pos=0.3] {\scriptsize$B\times T/4 \times d$} (dots2.south);
		\draw[arrow] ( dots2.north) --node[right,text=black!50] {} ([xshift=\x]  Encoder22.south);
		\draw[arrow,-] ( [xshift=\x] Encoder22.north) |- node[right,text=black!80,pos=0.2] {\scriptsize$B\times T/4 \times d$} (out_node2.south);
		
		\draw[|-,arrow] ( out_node2) |- node[right,text=black!50] {} (Decoder1.east);
		\draw[|-,arrow] ( out_node2) |- node[right,text=black!50] {} (Decoder2.east);

	}
	
	\node[above=of Encoder2,anchor=south, xshift=0cm,yshift=-0.02cm,align=right] () {\footnotesize \bf Magnitude Encoder};
	\node[above=of Encoder22,anchor=south, xshift=0cm,yshift=0.04cm,align=right] () {\footnotesize \bf Phase Encoder };
	\node[above=of Decoder2,anchor=south, xshift=0.14cm,yshift=0.00cm,align=right] () {\footnotesize\bf Fusion Decoder  };
	
	%\node[below=of input_character, xshift=0cm,yshift=0.8cm, text width=2cm, align=center] () {\small Previous \\ input tokens};
	%\node[below=of feats, xshift=0cm,yshift=1.0cm, text width=3cm, align=center] () {\small Magnitude \\ FBANK features};
	%\node[below=of feats2, xshift=0cm,yshift=1.0cm, text width=3cm, align=center] () {\small Phase \\ FBANK features};
	
	\begin{pgfonlayer}{background}
	\path (Encoder2.west |- Encoder2.north)+(-0.15,0.75) node (g) {};
	\path (Encoder1.east |- Encoder1.south)+(0.15,-0.2) node (h) {};
	\path[fill=green!25,rounded corners, draw=black!50,draw=none]
	(g) rectangle (h);
	\end{pgfonlayer}
	
	\begin{pgfonlayer}{background}
	\path (Encoder22.west |- Encoder22.north)+(-0.15,0.75) node (g) {};
	\path (Encoder12.east |- Encoder12.south)+(0.15,-0.2) node (h) {};
	\path[fill=green!25,rounded corners, draw=black!50,draw=none]
	(g) rectangle (h);
	\end{pgfonlayer}
	
	\begin{pgfonlayer}{background}
	\path (Decoder2.west |- Decoder2.north)+(-0.15,0.75) node (g) {};
	\path (Decoder1.east |- Decoder1.south)+(0.150,-0.2) node (h) {};
	\path[fill=blue!20,rounded corners, draw=black!50,draw=none]
	(g) rectangle (h);
	%\fill [line width=1mm,orange!10,rounded corners] (Decoder2.north  -| Decoder2.west) ++ (-0.9cm,+2.5cm) rectangle (Decoder1.south  -| Decoder1.east)  (+0.8cm,0.0cm);
	
	\end{pgfonlayer}
	\end{tikzpicture}\vspace{-3mm}
	\caption{\textbf{Transformer architecture} for the \textbf{middle fusion} approaches (\textsf{Fusion-Mid}) for training and inference, and for multi-encoder learning (\textsf{MEL}) during training. Two individual stream encoders are employed for each feature sequence; details of the decoder block are shown in Figure~\ref{fig:decoder_block}.
	}% 
	\label{fig:fusion_decoder}\vspace{-3mm}
\end{figure}
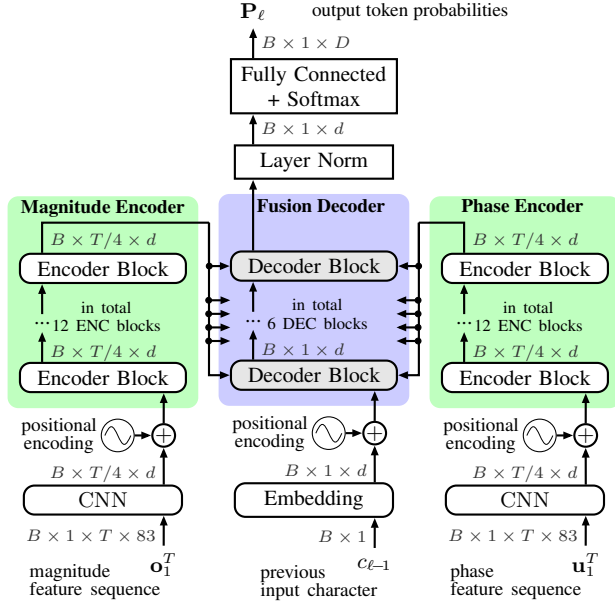

\section{Fusion Methods for End-to-End ASR}
\subsection{Early Fusion}\vspace{-1mm}
When it comes to fusion in end-to-end systems, the simplest approach is \textit{early fusion} as it is often applied in hybrid systems by stacking the individual feature vectors $\VEC{o}_t$ and $\VEC{u}_t$ to a joint feature representation $\VEC{x}_t=(\VEC{o}_t^\mathsf{T},\VEC{u}_t^\mathsf{T} )^\mathsf{T}$\!\!, with $(\ )^\mathsf{T}$ being the transposed. When using filterbank features, it has become a common technique to use convolutional neural networks (CNNs)~\cite{LeCun1998} in the input layer. In our \textsf{Fusion-Early} approach the additional feature stream is treated as an additional input channel, yielding an input tensor to the CNN block of size $\scriptsize B\!\times\!(2\cdot C)\!\times\!T\!\times\!F$ with $B$, $C$, $T$, $F$ being the batch size, channel depth, feature  sequence length, and feature dimension, respectively. After the input layer, the processing follows the standard transformer model architecture using a single attention-based encoder and a single decoder as in~\cite{Vaswani2017}.  

\subsection{Middle Fusion}
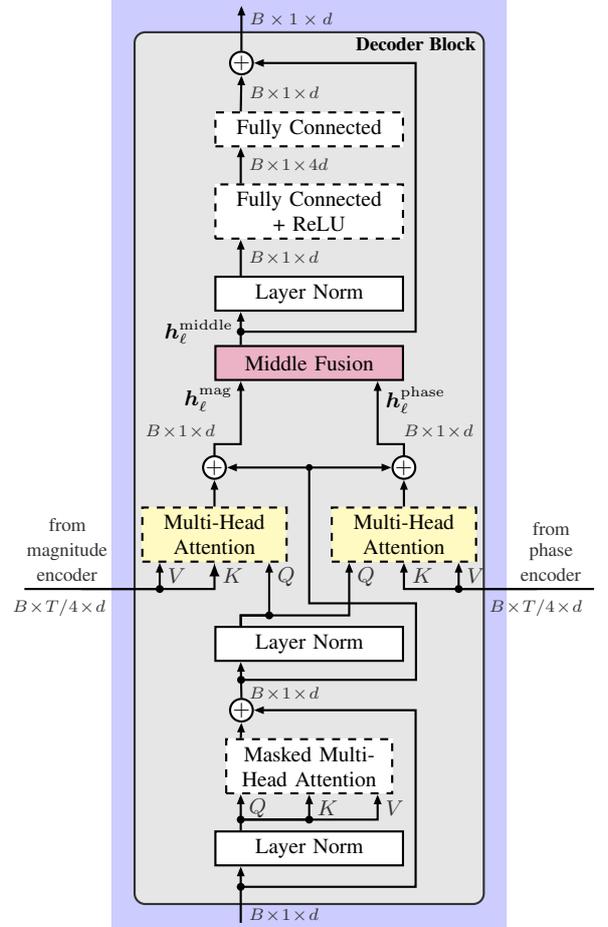
\begin{figure}[t]
	\centering
	\tikzset{%
		tipA/.tip={Triangle[angle=40:4pt]}
	}

	\tikzset{
		>=stealth',
		punktchain/.style={
			rectangle, 
			draw=black, thick,
			text width=8em, 
			minimum height=0.5cm, 
			text centered,fill=white, 
			on chain=encoder_block},
		line/.style={draw, thick, <-},
		arrow/.style={draw, thick,-tipA},
		loet/.style={draw,circle,fill=black,inner sep=0.3mm,outer sep=0.0mm},
		every join/.style={->, thick,shorten >=1pt},
	}
	
	\begin{tikzpicture}[
		node distance=0.75cm,
		mha/.style 		= {draw, fill=yellow!30,,text centered,text width=6em,thick}, 
		scale=1, every node/.style={scale=0.9}
		]

		{  [start chain=encoder_block going above]
			\def\x{-9mm}
			\def\sc{1.1}
			
			\node[punktchain] (ln1) {Layer Norm};
			\node[draw=none,below=of ln1,outer sep=0mm,inner sep=0mm,yshift=0.0cm] (encoder_input) {};
			
			\node[punktchain,dashed,yshift=-0.3cm,text width=7em]    (mha) {Masked Multi-Head Attention};
			
			\node[draw,circle,fill=white,on chain=encoder_block,thick,yshift=-0.6cm,xshift=\sc*\x,minimum size=0.1cm,inner sep=0mm]                            (res_in1) {$\mathbf{+}$};
			
			\node[punktchain,xshift=-\sc*\x,yshift=-0.3cm]        (ln_2) {Layer Norm};
			\node[on chain,yshift=5mm]          (tmp) {};
			
			\node[loet,on chain,yshift=-0.1cm] (tmp3) {};
			\node[punktchain,fill=purple!30,yshift=0.4cm] (combine) {Middle Fusion};

			%NODES ENC-DEC MHA
			\node[left=of tmp,mha,yshift=-0.1cm,dashed,xshift=0.7*-\x] (mha_1) {Multi-Head Attention};
			\node[right=of tmp,mha,yshift=-0.1cm,dashed, xshift=0.7*+\x] (mha_2) {Multi-Head Attention};

			\gettikzxy{(mha_1)}{\ax}{\ay}
			\gettikzxy{(tmp3)}{\bx}{\by}

			\node[draw,circle,fill=white,thick,minimum size=0.1cm,yshift=-0.0cm,inner sep=0mm] at (\ax,\by) (mha_resin_1) {$\mathbf{+}$};
			\gettikzxy{(mha_2)}{\cx}{\cy}
			\node[draw,circle,fill=white,thick,minimum size=0.1cm,yshift=-0.0cm,inner sep=0mm] at (\cx,\by)(mha_resin_2) {$\mathbf{+}$};
			
			%Fully Connected part
			\node[punktchain,yshift=-0.34cm] (ln_3) {Layer Norm};
			\node[punktchain,dashed,yshift=-0.3cm] (fc_1) {Fully Connected\\ + ReLU};
			\node[punktchain,dashed,yshift=-0.3cm] (fc_2) {Fully Connected};
			\node[draw,circle,fill=white,on chain=encoder_block,thick,yshift=-0.3cm,xshift=\sc*\x,minimum size=0.1cm,inner sep=0mm] (res_in3) {$\mathbf{+}$};

			%%% EDGES
			\draw[arrow] ( [xshift=\x]encoder_input.north) --node[right,text=black!80,yshift=-0.3cm] {\scriptsize$B\!\times\!1\!\times\!d$} ([xshift=\x] ln1.south);
			
			% first MHA input edges
			\draw[arrow] ([xshift=\x] ln1.north) --node[right,text=black!80] {} ([xshift=\x]mha.south)node[right,text=black!80,yshift=-0.2cm] {$Q$}; 
			\draw[arrow] ($ ([xshift=\x] ln1.north) !.3! ([xshift=\x]mha.south) $)node[loet]{} -| node[loet,pos=0.5]{} (mha.south) node[right,text=black!80,yshift=-0.2cm] {$K$};   
			\draw[arrow] ($ ([xshift=\x] ln1.north) !.3! ([xshift=\x]mha.south) $) -|  ([xshift=-\x]mha.south) node[right,text=black!80,yshift=-0.2cm] {$V$};  
			
			\draw[arrow,] ([xshift=\x] mha.north) --node[right,text=black!80,yshift=-0.1cm] {} (res_in1.south);	
			
			\draw[arrow,] (res_in1.north) --node[right,text=black!80,yshift=-0.2cm] {\scriptsize$B\!\times\!1\!\times\!d$} ([xshift=\x] ln_2.south);

			%%------ ENC1 MHA input edges
			% QUERY
			%\draw[arrow,] ([xshift=0.58*\x]ln_2.north) -- node[pos=0.79,right,text=black!80] {$Q$} ([xshift=-0.8*\x]mha_1.south);
			
			% KEY
			\draw[arrow,latex-] (mha_1.south) |- node[pos=0.25,right,text=black!80] {$K$} ([xshift=-1.75cm,yshift=-0.35cm]mha_1.south);
			\node[below=of mha_1,text=black!80, text width=2cm,xshift=-2.15cm, yshift=1.6cm, align=center] {\footnotesize from \\ magnitude encoder};
			\node[below=of mha_1,text=black!80, text width=1.7cm,xshift=-2.1cm,yshift=0.4cm] {\scriptsize$B\!\times\!T/4\!\times\!d$};
			
			% VALUE
			\draw[arrow,tipA-] ([xshift=0.8*\x]mha_1.south) |- node[loet,pos=0.5]{} node[pos=0.25,right,text=black!80] {$V$} ([xshift=-2.5cm,yshift=-0.35cm]mha_1.south);

			%%----- ENC2 MHA input edges
			% QUERY
			\draw[arrow] ([xshift=\x]ln_2.north) |- ([xshift=0.8*\x,yshift=-0.7cm]mha_2.south) --node[pos=0.74,right,text=black!80] {$Q$}  ([xshift=0.8*\x]mha_2.south);
			\draw[arrow] ([xshift=\x]ln_2.north) |- ([xshift=0.8*-\x,yshift=-0.7cm]mha_1.south) node[loet]{} -- node[pos=0.74,right,text=black!80] {$Q$}  ([xshift=0.8*-\x]mha_1.south);
			
			\draw[arrow,tipA-] (mha_2.south) |- node[pos=0.25,right,text=black!80] {$K$} ([xshift=1.75cm,yshift=-0.35cm]mha_2.south);
			\node[below=of mha_2,text=black!80, text width=2cm,xshift=2.15cm, yshift=1.55cm,align=center] {\footnotesize from \\ phase \\ encoder};
			\node[below=of mha_2,text=black!80, text width=1.7cm,xshift=2.1cm,yshift=0.4cm] {\scriptsize$B\!\times\!T/4\!\times\!d$};
			
			\draw[arrow,tipA-] ([xshift=-0.8*\x]mha_2.south) |- node[loet,pos=0.5]{} node[pos=0.25,right,text=black!80] {$V$} ([xshift=2.5cm,yshift=-0.35cm]mha_2.south);

			% Residual Connections
			\draw[arrow] ($ ([xshift=\x]encoder_input.south) !.5! ([xshift=\x]ln1) $)node[loet]{} -| ++(2.3,00) |- (res_in1);
			
			% "The crooked one"
			\draw[arrow,-] ($ (res_in1.north) !.5! ([xshift=\x]ln_2.south) $)node[loet]{} -| ++(2.3,00) |- ([yshift=3mm]ln_2.north) -- (tmp3);
			
			\draw[arrow] ($ ([xshift=\x]combine.north) !.4! ([xshift=\x]ln_3.south) $)node[loet]{} -| ++(2.3,00) |- (res_in3);

			\draw[arrow] (tmp3) -- (mha_resin_1); 
			\draw[arrow] (tmp3) -- (mha_resin_2);

			\draw[arrow] (mha_1) -- (mha_resin_1); 
			\draw[arrow] (mha_2) -- (mha_resin_2);

			\draw[arrow] ([xshift=\x] combine.north) --node[left,text=black,pos=0.4] {$\VECG{h}_\ell^{\mathrm{middle}}$} ([xshift=\x] ln_3.south);	
			
			\draw[arrow] ([xshift=\x] ln_3.north) --node[right,text=black!80,pos=0.5] {\scriptsize$B\!\times\!1\!\times\!d$} ([xshift=\x] fc_1.south);
			
			\draw[arrow] ([xshift=\x] fc_1.north) --node[right,text=black!80,pos=0.5] {\scriptsize$B\!\times\!1\!\times\!4d$} ([xshift=\x] fc_2.south);
			
			\draw[arrow] ([xshift=\x] fc_2.north) --node[right,text=black!80,pos=0.5] {\scriptsize$B\!\times\!1\!\times\!d$} (res_in3.south);
			
			\draw[arrow] ( res_in3.north) --node[pos=0.7,right,text=black!80] {\scriptsize$B\times 1 \times d$} +(0,0.6cm);
			
			\draw[arrow] (mha_resin_1.north) |-node[pos=0.4,above,text=black!80, xshift=-0.5cm] {\scriptsize$B\!\times\!1\!\times\!d$} ($ (mha_resin_1.north) !.15! ([xshift=\x]combine.south) $) -| node[pos=0.85,left ,text=black!100,yshift=-0.0cm,xshift=0.0cm] {$\VECG{h}_\ell^{\mathrm{mag}}$} ([xshift=\x]combine.south) ;
			\draw[arrow] (mha_resin_2.north) |-node[pos=0.4,above ,text=black!80,yshift=-0.0cm,xshift=0.5cm] {\scriptsize$B\!\times\!1\!\times\!d$} ($ (mha_resin_2.north) !.15! ([xshift=-\x]combine.south) $) -| node[pos=0.85,right ,text=black!100,yshift=-0.0cm,xshift=0.0cm] {$\VECG{h}_\ell^{\mathrm{phase}}$} ([xshift=-\x]combine.south);

			\node[right=of res_in3,anchor=east, xshift=2.6cm,yshift=0.28cm,align=right] () {\footnotesize \bf Decoder Block };
		}

		\begin{pgfonlayer}{background}
			\path (fc_2.west |- fc_2.north)+(-1.35,1.5) node (g) {};
			\path (ln1.east |- ln1.south)+(1.35,-0.8) node (h) {};
			\path[fill=blue!20,]
			(g) rectangle (h);
		\end{pgfonlayer}
		
		\begin{pgfonlayer}{background}
			\path (fc_2.west |- fc_2.north)+(-1.05,1.05) node (g) {};
			\path (ln1.east |- ln1.south)+(1.05,-0.5) node (h) {};
			\path[fill=black!10,rounded corners,draw=black!80, thick]
			(g) rectangle (h);
		\end{pgfonlayer}

	\end{tikzpicture}\vspace{-3mm}
	\caption{Single \textbf{decoder block} (cf.\ Figure~\ref{fig:fusion_decoder}) for the \textbf{middle fusion} approach  of both encoder-decoder multi-head attention block outputs. The same setup is used for the \textbf{Multi-Encoder Learning} approaches (\textsf{MEL-t-mag} and \textsf{MEL-t-phase}) only during training, while only one encoder is active in inference. Dropout layers~\cite{Nitish2014} are in dashed line boxes. 
	}% 
	\label{fig:decoder_block}\vspace{-3mm}
		\centering
\end{figure}

\vspace{-1mm}
For the \textit{middle fusion} approaches we use two individual stream encoders for each feature sequence as shown as the green boxes in Figure~\ref{fig:fusion_decoder}. Based on the previous output token $c_{\ell-1}\in\mathcal{C}=\{c^{(1)},c^{(2)},\dots,c^{(D)}\}$ and the entire feature sequences $\VEC{o}_1^T$ and $\VEC{u}_1^T$, where $t\!\in\!\{1,\dots,T\}$ and $\ell\!\in\!\{1,\dots,L\}$ are time instants of the input feature vector and output token sequences, respectively, the transformer outputs a vector $\VEC{P}_\ell$ with output token probabilities for the current sequence time instant $\ell$. Each of the stream encoders comprises a total of 12 identical encoder blocks, each consisting of the multi-head self-attention mechanism and position-wise fully connected layers as in~\cite{Vaswani2017}. The output of the last encoder block is then passed on to each of the in total 6 decoder blocks, which are detailed in Figure~\ref{fig:decoder_block}. We investigate two different strategies for middle fusion, both using two separate encoder-decoder multi-head attention blocks (shown in yellow) for each stream encoder. Fusion is then applied to the hidden entities $\VECG{h}_\ell^{\mathrm{mag}}$ and $\VECG{h}_\ell^{\mathrm{phase}}$ after the two encoder-decoder multi-head attention blocks for each stream, shown as red block in Figure~\ref{fig:decoder_block}, yielding $\VECG{h}_\ell^{\mathrm{middle}}$. First is the weighted sum approach, dubbed \textsf{Fusion-Mid-WS}, employing a simple linear combination
\begin{equation}\vspace{-1mm}
\VEC{h}_\ell^{\mathrm{middle}} = \alpha\VECG{h}_\ell^{\mathrm{mag}} + (1-\alpha)\VECG{h}_\ell^{\mathrm{phase}}\vspace{-1mm}
\end{equation} 
with $\alpha\!\in\![0,1]$ being a fusion weight to balance the influence of each of the encoder-decoder multi-head attention blocks. In addition, for the \textsf{Fusion-t-Mid-WS} approach, we tied ("-\textsf{t}-") the parameters of both involved encoder-decoder multi-head attention blocks. 

The second variant dubbed \textsf{Fusion-Mid-CC} is the straightforward concatenation of both entities according to $\VEC{h}_\ell^{\mathrm{middle}} = ((\VECG{h}_\ell^{\mathrm{mag}})^\mathsf{T},(\VECG{h}_\ell^{\mathrm{phase}})^\mathsf{T})^\mathsf{T}$ as it has been used for audiovisual speech recognition in~\cite{Afouras2018}. To still allow residual connections, in this case it becomes necessary to halve the dimension of both encoder-decoder multi-head attention block outputs to $d/2$ and add the residual from the self-attention after the concatenation, where the previous model dimension is restored.

\newcommand{\f}[1]{\textbf{#1}}
\newcommand{\s}[1]{\underline{#1}}
\renewcommand{\tabcolsep}{0.2cm}

\subsection{Late Fusion}\vspace{-1mm}
As \textit{late fusion} we investigate the fusion of output token probability vectors $\VEC{P}_\ell^{\mathrm{mag}}$ and $\VEC{P}_\ell^{\mathrm{phase}}$ stemming from separately trained transformer networks for each feature stream $\VEC{o}_1^T$ and $\VEC{u}_1^T$. The final output token probability in the log domain for each time instant $\ell$ is then computed as (\textsf{Fusion-Late})\vspace{-4mm}

\begin{equation}\vspace{-1mm}
   \log\VEC{P}_\ell^{\mathrm{late}} = \beta\log\VEC{P}_\ell^{\mathrm{mag}} + (1-\beta)\log\VEC{P}_\ell^{\mathrm{phase}}
\end{equation} 
with $\beta\in[0,1] $ being a posterior fusion weight and $\log(\ )$ \mbox{operating} element-wise. One major advantage of the late fusion approach is that it uses independently trained models, and the balancing hyperparameter $\beta$ can be easily set during inference time if one feature stream deteriorates.
\vspace{-1mm}
\subsection{Novel Multi-Encoder Learning (MEL)} \vspace{-1mm}
In addition to the previous fusion paradigms, we employ a novel yet simple multi-encoder learning (MEL) approach to investigate if the additional information during training helps to increase robustness without using any additional parameters in inference. For this method, we train the middle fusion transformer model exactly as for the \textsf{Fusion-t-Mid-WS} approach using both encoders, but tie ("\textsf{-t-}") all parameters of both multi-head attention blocks (shown as yellow blocks in Figure~\ref{fig:fusion_decoder}). During inference, however, only \textit{one} of the encoders is active and the decoder uses one instance of the jointly trained multi-head attention. For the \textsf{MEL-t-mag} and \textsf{MEL-t-phase} approaches, only the magnitude or the phase encoder is active during inference, respectively, while the fusion weight $\alpha=0.9$ \textit{during training} is biased towards the inference encoder. Both models trained with the MEL method can also be subject to late fusion, dubbed \textsf{MEL-t-Fusion-Late} in Tables~\ref{tab:wsj_lm} and~\ref{tab:libri}.

%\begin{table}[t!]
%	\centering
%	
%	\begin{tabular}{p{2.4cm} @{\hskip 0.25cm}m{0.7cm} c c c c}
%		\toprule
%		\multirow{3}*{Approach} & \multirow{3}{1cm}{ \# of \newline param. \newline inference} & \multicolumn{2}{c}{dev93} & \multicolumn{2}{c}{eval92} \\
%		\cmidrule(lr){3-4} \cmidrule(lr){5-6} & & WER & CER & WER & CER \\
%		& & &  &  &  \\
%		\midrule
%		\textsf{Baseline-mag}   & 16.8M 	& 14.62 	& 5.28 		& 11.66  	& 4.03 		\\
%		\textsf{Baseline-phase} & 16.8M 	& 15.75 	& 5.79  	& 12.90  	& 4.33 		\\
%		\midrule
%		\textsf{Fusion-Early} 	& 16.8M 	& 14.46 	& 5.07 		& 10.83 	& 3.70 		\\
%		\textsf{Fusion-Mid-CC} 	& 28.4M  	& 14.80 	& 5.26 		& 11.31  	& 3.81 		\\
%		\textsf{Fusion-Mid-WS} 	& 27.2M 	& 16.82 	& 5.75 		& 13.43  	& 4.11 		\\
%		\textsf{Fusion-t-Mid-WS} 	& 27.2M & 14.78 	& 5.10 		& 11.57  	& 3.80 		\\
%		\textsf{Fusion-Late}    & 33.5M 	& \f{13.38}	& \f{4.79}  & \f{10.65}	& \f{3.53} 	\\
%		\midrule
%		\textsf{MEL-t-mag}		& 16.8M 	& 14.91 	& 5.09 		& 11.48 	& 3.66 		\\
%		\textsf{MEL-t-phase}	& 16.8M 	& 16.01 	& 5.85 		& 12.09 	& 4.25 		\\
%		\textsf{MEL-t-Fusion}	& 16.8M 	& ?? 		& ?? 		& ?? 		& ?? 		\\
%		\bottomrule
%	\end{tabular}
%	\caption{Transformer-based approaches \textbf{w/o language model} on the \textbf{Wall Street Journal} task}
%	\label{tab:wsj_nolm}
%	\vspace{-8mm}
%\end{table}

\vspace{-1mm}
\subsection{Language  Model and Decoding}\vspace{-1mm}
For all investigated approaches including non-fusion baselines \textsf{Baseline-mag} and \textsf{Baseline-phase}, we use beam-search decoding during inference and slightly deviate from the standard transformer architecture in~\cite{Vaswani2017} by using layer normalization before each attention or stack of fully connected layers according to the implementation in~\cite{Vaswani2018}. During decoding, the final output $\mathbf{P}_\ell^\mathrm{final}$ of \textit{all} approaches can optionally be computed as 
\begin{equation}\vspace{0mm}
	\log\VEC{P}_\ell^{\mathrm{final}} = \log\VEC{P}_\ell  + \lambda\log\VEC{P}_\ell^{\mathrm{LM}},\vspace{0mm}
\end{equation} 
adding logarithmic character probabilities from the language model $\VEC{P}_\ell^{\mathrm{LM}}$, with the standard language model weight $\lambda$ chosen according to the shallow integration technique~\cite{gulcehre2015}. For experiments on the Wall Street Journal task we report all results both without and with additional language model in Table~\ref{tab:wsj_lm}.  
\vspace{-1mm}
\section{Experimental Setup}\vspace{-1mm}

\subsection{Databases}	\vspace{-1mm}
We evaluate our fusion approaches on the 81-hour Wall Street Journal (WSJ) dataset~\cite{paul1992} using the \texttt{dev93} and \texttt{eval92} splits to evaluate system performance in terms of word error rate $\mathrm{WER}=1-\frac{N-D-I-S}{N}$, as well as w.r.t.\ character error rate (CER), where the number of units $N$, deletions $D$, insertions $I$, and substitutions $S$ are calculated on character-level instead of on word-level as for the WER. To investigate our approaches also on a large-scale dataset, all experiments are repeated on Librispeech~\cite{Panayotov2015} using the 960\,h training set along with the \texttt{clean} and \texttt{other} portions of the dev and test datasets. All used speech signals are sampled at 16\,kHz and analyzed with a 25\,ms window and a frame shift of 10\,ms. 
\vspace{-1mm}
\subsection{Acoustic Frontends}\vspace{-1mm}
For the middle fusion approaches, each of the encoders receives a sequence of $T$ feature vectors of dimension $F=83$. As magnitude features $\VEC{o}_1^T$ we use standard 80-dimensional filterbank features extended with 3-dimensional pitch features extracted with the \texttt{Kaldi} toolkit~\cite{Povey2011}. For the phase features $\VEC{u}_1^T$ we follow the processing of~\cite{loweimi2013,Lohrenz2018} and use the group-delay representation extracted from an all-pole model, and also apply an 80-dimensional mel-filterbank. For details on the processing, please refer to~\cite{Lohrenz2018}. 
The convolutional neural networks (CNNs) at the input layers, shown as CNN blocks in Figure~\ref{fig:fusion_decoder}, consist of a total of four convolutional layers each using $3\!\times\!3$ filter kernels. The second and forth convolutional layer use a stride of 2 in both temporal and frequency direction thus compressing the input sequence length to $T/4$. We note that it might be beneficial to apply separate convolutions to the pitch features but follow~\cite{wang2019e,moriya2020} for comparability.

\vspace{-1mm}
\subsection{Acoustic and Language Model Configuration}\vspace{-1mm}
As shown in Figure~\ref{fig:fusion_decoder}, the used transformer architecture for the acoustic model follows the standard architecture from~\cite{Vaswani2017} employing a total of 12 encoder blocks for each used encoder, while the decoder stacks 6 decoder blocks. For WSJ, the model dimension is set to $d=256$ and multi-head attention blocks use 4 attention heads, while for Librispeech we use a larger model, where both values are doubled to $d=512$ and 8 attention heads. Transformer models were trained using the Adam optimizer with label-smoothed cross-entropy loss~\cite{mueller2020}. We follow~\cite{Dong2018c} for learning rate scheduling. For Librispeech experiments we additionally used spectral augmentation~\cite{Park2019d}.

For language modeling in WSJ experiments, we apply a 3-layer LSTM network with a size of 1200 each, which is trained on word-level but yields character-level probabilities $\VEC{P}_\ell^{\mathrm{LM}}$ for a total of $D\!=\!52$ characters, using the lookahead method proposed in ~\cite{hori2018}. As language model weight we follow~\cite{wang2019e} and choose $\lambda=0.9$. For Librispeech we use \texttt{SentencePiece} for word tokenization with an output token dimension size of $D=5000$ embeddings~\cite{Kudo2018} and use a 4-layer LSTM as token-based LM with each layer having a size of 1024. The language model weight for Librispeech is set to $\lambda=0.4$ following~\cite{wang2019e}.

For the \textsf{Fusion-Late} approach, the posterior fusion weight $\beta$ (only applied during inference) has been optimized on the respective development sets. For all middle fusion approaches with weighted sum we set the fusion weight $\alpha\!=\!0.9$ without further tuning. The same value was used for the MEL approaches during training for the respective primary encoder (magnitude encoder for \textsf{MEL-t-mag} and phase encoder for \textsf{MEL-t-phase}). 

All models were trained using the \texttt{espresso} and \texttt{fairseq} toolkits based on \texttt{PyTorch}~\cite{wang2019e,ott2019,Paszke2019}. WSJ models were trained on a single GTX1080Ti GPU, while Librispeech models used 4 Tesla P100 GPUs. All experiments use the same random seed.

\renewcommand{\arraystretch}{0.95}

%%%%%%%%% HEADER%%%%%%%%%%

\begin{table*}[t!]
	\centering
	
	\begin{tabular}{@{\hskip 0.1cm} p{2.9cm} c @{\hskip 0.15cm} c @{\hskip 0.2cm} c c c c c c @{\hskip 0.3cm} c c c c @{\hskip 0.05cm} }
		\toprule
		\multirow{4}{*}{Approach} & \multicolumn{2}{c}{\multirow{2}{*}{\shortstack{Inference complexity }}}  & & \multicolumn{4}{c}{\textit{Without} language model} & & \multicolumn{4}{c}{\textit{With} language model} \\
		& & & & \multicolumn{2}{c}{\multirow{2}{*}{\tt{dev93}}}  & \multicolumn{2}{c}{\multirow{2}{*}{\tt{eval92}}} & & \multicolumn{2}{c}{\multirow{2}{*}{\tt{dev93}}} \vspace{-2mm} & \multicolumn{2}{c}{\multirow{2}{*}{\tt{eval92}}}\\
		
		& \multirow{3}{*}{\shortstack{\# of \\ parameters}} & \multirow{3}{*}{\shortstack{relative \\ runtime}} & & & & & & & & & & \\
		
		\cmidrule(lr){5-6} \cmidrule(lr){7-8} \cmidrule(lr){10-11} \cmidrule(lr){12-13}
		& 		& 																	& & WER 			& CER 		& WER 		& CER 		&	& WER 		& CER 		& WER 		& CER		\\
		\midrule
		\textsf{Baseline-mag}    	& \multirow{2}{*}{16.8M}& \multirow{2}{*}{1.0}	& & 14.62 		& 5.28 		& 11.66  	& 4.03 	 	&	& 6.51 		& 3.64 		& 4.43  	& 2.37 		\\
		\textsf{Baseline-phase} 	& 	 					&						& & 15.75 		& 5.79  	& 12.90  	& 4.33 	 	&	& 7.32 		& 4.23 		& 5.48  	& 3.17 		\\
		\midrule                          		
		\textsf{Fusion-Early} 		& 16.8M 				& 1.02					& & 14.46 		& 5.07 		& 10.83 	& 3.70 	 	&	& 6.57		& 3.39 		& 4.41 		& 2.27 		\\
		\textsf{Fusion-Mid-CC} 		& 28.4M 				& 1.36					& & 17.11 		& 6.17 		& 11.31  	& 3.81 	 	&	& 6.58 		& 3.71 		& 4.20 		& 2.23 		\\
		\textsf{Fusion-Mid-WS} 		& 28.8M 				& 1.37					& & 16.82 		& 5.75 		& 13.43  	& 4.11 	 	&	& 6.40 		& 3.55 		& 4.38  	& 2.46 		\\
		\textsf{Fusion-t-Mid-WS} 	& 27.2M 				& 1.33 					& & 15.89 		& 5.58 		& 12.28  	& 4.07	 	&	& 6.23		& 3.57 		& \s{4.09} 		& \s{2.10} 		\\
		\textsf{Fusion-Late}    	& 33.5M 				& 1.81  				& & \f{13.38}	& \f{4.79}  & \s{10.65}	& \s{3.53} 	&	& \s{5.79} 		& \s{3.08}  & 4.31  	& 2.32 		\\
		\midrule	                      
		\textsf{MEL-t-mag}			& 16.8M 				& \multirow{2}{*}{1.0}	& & 15.22 		& 5.36 		& 11.73 	& 3.95 	 	&	& 6.29 	& 3.43 		&  4.31 &  2.50 \\
		\textsf{MEL-t-phase}		& 16.8M 				& 						& & 16.01 		& 5.85 		& 12.09 	& 4.25 	 	&	& 7.00 		& 4.04 		&  4.68  	&  2.74 	\\
		\textsf{MEL-t-Fusion-Late}	& 33.5M 				& 1.81					& & \s{14.04} 	& \s{4.90} 	& \f{10.12}	& \f{3.38} 	&	& \f{5.50} 	& \f{3.01} 	& \f{3.40} 	& \f{1.95} 	\\
		\bottomrule
	\end{tabular}
	\caption{Transformer-based approaches on the \textbf{WSJ} task. Best results are \f{bold}, second best results are \s{underlined}.}
	\label{tab:wsj_lm}
	\vspace{-9mm}
\end{table*}

\begin{table}[t]
	\centering
	\begin{tabular}{l c c}
		\toprule
		Approach & \tt dev93 & \tt eval92 \\
		\midrule
		%Human 2015 								 	& 8.08 	& 5.03 \\
		%Hori et al. 2018 ~\cite{hori2018} 		  	& 8.4 	& 5.1 \\
		Tsunoo et al. 2019~\cite{Tsunoo2019} 	  & - 	& 5.00 \\ 
		Karita et al. 2019~\cite{Karita2019b} 	  & 7.70 	& 4.50 \\
		Karita et al. 2019~\cite{Karita2019}	  & 6.80 	& 4.40 \\
		Moriya et al. 2020~\cite{moriya2020} 	 & 6.90 	& 4.20 \\
		\midrule
		Ours  & \textbf{5.50} & \textbf{3.40} \\
		\bottomrule
	\end{tabular}
	\caption{WER comparison of recent transformer-based end-to-end ASR approaches \textbf{with language model} on \textbf{WSJ}.}
	\label{tab:sota}
	\vspace{-9mm}
\end{table} 

\vspace{-2mm}
\section{Recognition Results and Discussion}\vspace{-1mm}
% % Wallstreet Experiments
Results of all approaches on the WSJ task are shown in Table~\ref{tab:wsj_lm}. For the single-encoder approaches, we note that the \textsf{Baseline-mag} transformer performs slightly better than the \textsf{Baseline-phase} approach (about 1\% absolute in terms of WER on the \texttt{eval92} set with language model). Comparing results \textit{without} and \textit{with} language model (LM) for both \textsf{Baseline} approaches, we note that with LM the error rates are significantly reduced (especially for word errors), showing the effectiveness of the word-based lookahead LM~\cite{hori2018}. 

Considering all \textsf{Fusion-X} approaches, an overall improvement compared to \textsf{Baseline-Y} methods is visible in most cases, suggesting that the phase-based speech representation indeed yields complementary information that can improve recognition. Interestingly, none of the \textsf{Fusion-Mid} approaches is able to provide consistent improvements. While only slightly increasing the size and complexity of the transformer, the common \textsf{Fusion-Early} approach is not able to decrease the WER with language model on the \texttt{dev93} set. Among the \textsf{Fusion-Mid} approaches with LM, the \textsf{Fusion-t-Mid-WS} variant performs best by achieving 4.09\% WER on \texttt{eval92} while also adding the least complexity to the model during inference. The fully modular \textsf{Fusion-Late} approach yields the highest computational complexity as both baseline transformers have to be inferred. On the other hand, it offers an inference-time parameter $\beta$ to balance the fusion and performs remarkably well \textit{without} LM. \textit{With} LM, however, \textsf{Fusion-Late} is not able to generalize the superior \texttt{dev93} performance towards the \texttt{eval92} set.    

Our novel \textsf{MEL-t-mag} and \textsf{MEL-t-phase} approaches with LM achieve a WER reduction on \texttt{eval92} of 0.12\% and even 0.8\% \textit{absolute} compared to their respective \textsf{Baseline} approaches without adding any additional complexity during inference. This suggests that the parameter tying of both encoder-decoder multi-head attention blocks in the course of our multi-encoder learning (MEL) strongly improves robustness and generalization, especially for the mid-size WSJ training set. Combining both improved MEL-based models in the~\textsf{MEL-t-Fusion-Late} approach yields the lowest WER of 3.40\% on \texttt{eval92}, corresponding to a WER reduction of 0.91\% absolute w.r.t.\ the normal \textsf{Fusion-Late} approach (4.31\%), and a remarkable reduction of up to 19\% relative compared to the best recently published transformer-based approach by Moriya et al.~\cite{moriya2020}, as shown in Table~\ref{tab:sota}.  

\renewcommand{\tabcolsep}{0.015cm}
\begin{table}[t!]
	\centering
	\begin{tabular}{@{\hskip 0cm} p{2.85cm}@{\hskip -0.2cm} c c c c c @{\hskip -0.15cm}}
		\toprule
		\multirow{3}{*}{Approach} & \multirow{3}{*}{ \shortstack{\# of \vspace{-0.6mm}\\ inference \\ param.}} & \multicolumn{4}{c}{\multirow{1}{*}{\hspace{-1mm}WER}}  \\
		&  & {\hskip -0.1cm} \multirow{2}{1cm}{\tt\footnotesize \hspace{0.5mm}\shortstack{ dev\\clean}} & \multirow{2}{1cm}{\tt \hspace{0.5mm}\footnotesize \shortstack{dev \\ other}} & \multirow{2}{1cm}{\tt \footnotesize \hspace{0.5mm}\shortstack{test\\clean}} & \multirow{2}{1cm}{\tt \footnotesize \hspace{0.5mm}\shortstack{test\\other}}\\ 
		& & &  &  &  \\
		\midrule
		\textsf{Baseline-mag}    	& 69.8M 	& 3.44 	& 7.80 	& 4.05 	& 8.14 \\
		\textsf{Baseline-phase} 	& 69.8M 	& 3.80 	& 9.00 	& 4.43  & 9.62 \\
		\midrule
		\textsf{Fusion-Early} 		& 69.8M 	& 3.35 	& 7.82  & 3.78 	& 8.14 \\
		\textsf{Fusion-Mid-CC} 		& 114.0M 	& 3.70 	& 7.98 	& 4.08  & 8.25 \\
		\textsf{Fusion-Mid-WS} 		& 115.6M	& 3.43	& 8.02 	& 3.96  & 8.34 \\
		\textsf{Fusion-t-Mid-WS} 	& 109.3M 	& 3.35  & 7.26 	& 3.77 	& 7.68 \\
		\textsf{Fusion-Late}     	& 139.6M 	& \f{2.99} 	& 6.91 	& 3.63 	& 7.34 \\
		\midrule
		\textsf{MEL-t-mag} 			& 69.8M 	& 3.37 	& 7.68 	& 3.87  & 7.90 \\
		\textsf{MEL-t-phase} 		& 69.8M 	& 3.68  & 8.66  & 4.05  & 9.03   \\
		\textsf{MEL-t-Fusion-Late} 	& 139.6M 	& 3.05 	& \f{6.63} 	& \f{3.34}  & \f{7.15} \\
		\bottomrule
	\end{tabular}
	\caption{Transformer-based approaches on \textbf{Librispeech}.}
	\label{tab:libri}
	\vspace{-8mm}
\end{table}

All approaches are also evaluated on the Librispeech task with results being reported in Table~\ref{tab:libri}. Among the fusion approaches only \textsf{Fusion-t-Mid-WS} and \textsf{Fusion-Late} yield consistent improvement over all data splits compared to both baselines, while \textsf{Fusion-Late} is the better yet more costly one. As for the WSJ task, also on Librispeech our \textsf{MEL-t-X} approaches consistently outperform the respective \textsf{Baseline} approaches (4.4\% and 8.6\% relative improvement on \texttt{test\,clean}, respectively), while having equal inference complexity. With late fusion of both MEL-enhanced transformer models (\textsf{MEL-t-Fusion-Late}), we achieve our best results on Librispeech with a remarkable WER reduction of 17.5\% and 12.2\% relative on the \texttt{clean} and \texttt{other} portions of the test set, respectively, compared to the standard transformer approach (\textsf{Baseline-mag}).
\vspace{-1mm}
\section{Conclusion}\vspace{-1mm}
In this contribution we introduced several fusion mechanisms to transformer-based end-to-end speech recognition. In addition, we apply a novel multi-encoder learning method (MEL), that uses the additional information from a second encoder only during training, while just a single encoder is used during inference. Compared to standard transformer approaches our novel MEL achieves a consistent WER reduction on all investigated tasks at the same runtime and number of parameters. By performing additional fusion, we achieve a WER reduction of 19\% relative on the Wall Street Journal task compared to state of the art, thereby defining a new benchmark for transformer-based ASR on that task.
\vspace{-3mm}

\section{Acknowledgements}\vspace{-1mm}
The research leading to these results has received funding from the Deutsche Forschungsgemeinschaft (DFG, German Research Foundation) for project number 414091002, as well as from the Bundesministerium f\"ur Wirtschaft und Energie (BMWi) under funding code 01MK20011T.

\clearpage

\bibliographystyle{IEEEtran}

\bibliography{ifn_spaml_bibliography}
\end{document}

%% file: macros/tikz_init.tex
\usepackage{tikz}
\usepackage{pgfplots}

\pgfplotsset{compat=newest}

\usetikzlibrary{arrows,automata}
\usetikzlibrary{shapes.geometric,shapes.arrows,decorations.pathmorphing}
\usetikzlibrary{matrix,chains,scopes,positioning,arrows,fit}
\usetikzlibrary{patterns}
\usetikzlibrary{chains,backgrounds,calc}

%% file: macros/mathmacros.tex
\newcommand{\VEC}[1]{\mathbf{#1}}          % Vector
\newcommand{\VECG}[1]{\boldsymbol{#1}}     % Vector with Greek letter
\newcommand{\putindex}[3]{\vtop{\hbox{\hspace{#3} $#1$}
            \hbox{\raise 6mm \hbox{$\scriptscriptstyle #2$}}}}

\newcommand{\gradx}[0]{\vtop{\hbox{\rm grad}
            \hbox{\raise 2.5mm \hbox{\rm \hspace{2mm} \footnotesize x}}}}

\newcommand{\grady}[0]{\vtop{\hbox{\rm grad}
            \hbox{\raise 2.5mm \hbox{\rm \hspace{2mm} \footnotesize y}}}}

\newcommand{\grad}[1]{\vtop{\hbox{\rm grad}
            \hbox{\raise 2.5mm \hbox{#1}}}}

\newcommand{\btb}{     \begin{tabbing}             }
\newcommand{\bte}{     \end{tabbing}               }